\documentclass[a4paper,12pt]{article}

\usepackage{amsmath}
\usepackage{hyperref}
\usepackage{cite}
\usepackage{amssymb}
\usepackage{braket}
\title{Complete ADHM Sigma Model}
\author{Abbas Ali\footnote{Email: aali.ph@gmail.com} and P.P. Abdul Salih\footnote{Email: ppsalih@gmail.com}\\
		Physics Department, Aligarh Muslim University,\\ Aligarh, India }
\date{}
\begin{document}

\maketitle

\begin{abstract}

The field theoretic ADHM
instantons have stringy generalizations as  linear sigma models. These were constructed by Witten in 1995. Recently Ali and Ilahi constructed a complementary version  related to Witten's construction by a duality. In this note we generalize these to a complete ADHM instanton linear sigma model, as suggested by Witten, in which
above duality is promoted to a $Z_2$ symmetry.	

\end{abstract}

\section{Introduction}\label{intro}

The SU(2) Yang-Mills instanton by 't Hooft is a field theoretic construct that depends upon $5k+4$ parameters for topological charge $k$. This can be promoted  to string theory using sigma model approach. Corresponding worldsheet construction was done in Ref.\cite{Callan:1991dj} by Callan, Harvey and Strominger. It obeys  the low energy effective equations derived from heterotic strings. Corresponding solution possesses (0,4) supersymmetry. There is a  world-brane version too that was studied in Ref.\cite{Callan:1991ky}  while overall structure was reviewed in Ref.\cite{Callan:1991at}.

The full SU(2) Yang-Mills  instanton can have $8k$ parameters for topological charge $k$ and this was constructed  by Atiyah, Drinfeld, Hitchin and Manin in 1978 in a remarkably brief and linear algebraic construction in Ref.\cite{Atiyah:1978ri}. This construction used inspirations from twistor theory \cite{Witten:1978qe}. In Ref.\cite{Corrigan:1983sv} it was explained in terms of the differential geometry of $R^4$. Further accessible version and generalization to other groups were presented in Ref.\cite{Christ:1978jy} and \cite{Corrigan:1978ce}. A recent mathematical exposition of the construction is covered in Ref.\cite{Donaldson:2022}. 

Witten gave us an equally remarkable stringy generalization using linear, that is massive, sigma model in Ref.\cite{Witten:1994tz}. Insights related to renormalization and quantization of this ADHM instanton sigma model including some clarifcation of superfield structure were presented in Ref.\cite{Lambert:1995dp} by Lambert. The version to include D-branes was constructed by Douglas in Ref.\cite{Douglas:1996uz}. Related $p$-brane and $D$-brane aspects were further investigated in Refs.\cite{Lambert:1996yd,Lambert:1997gs}. The aspects related to the geometry of the semi-infinite throat \cite{Callan:1991at} were investigated by Johnson in Ref.\cite{Johnson:1998yw}. 

In spite of these deep and powerful insights the ADHM sigma model is not as well explored and exploited as the simpler version of the Callan, Harvey and Strominger for the 't Hooft case. This situation should change with the results of Ref.\cite{Ali:2023csc} where an ADHM sigma model complementary to the one in Ref.\cite{Witten:1994tz} was constructed by Ali and Ilahi (see also \cite{Ali:2023kkf, Ali:2023xov}). The two models, Witten's original and Ali-Ilahi's complementary model, are related to each other by a simple duality. A version of this duality was already perceptible  in Ref.\cite{Lambert:1997gs}. The construction in Ref. \cite{Ali:2023csc} resolves the mystery that was recorded in \cite{Witten:1994tz} and was further analyzed in Ref.\cite{Witten:1995zh} without mitigation. Its quantization has been discussed in Ref.\cite{Ali:2023zxt}. It has already helped us in resolving the long standing problem of doubling of the Ramond superalgebra encountered in the well known study of $AdS_3$ superstrings in Ref.\cite{Giveon:1998ns}. This in turn is related to the BTZ black hole \cite{Banados:1992wn} and its stringy generalization \cite{Ali:1992mj}.

The present note contains the construction of a complete ADHM sigma model as well as discussion of related issues. It is complete in the sense that this construction incorporates the duality between the original and complementary  model as a $Z_2$ symmetry.

The rest of this article is organized as follows. In Section \ref{nbis} we spell out the notation, set out the background,  collect ingredients from the Original and Complementary ADHM instanton sigma models and combine these to synthesize the complete model. It has a $k+k'$ instanton structure, that is, it possesses two topological charges. The $1+1$  instanton case, that is $k= k'=1$,  is taken up in Section \ref{one} to discuss the structure of resulting moduli space. We discuss the other aspects of the resulting model in Section \ref{discussion}.
 
\section{Notation, Background, Ingredients, Synthesis}\label{nbis}

We begin by collecting in this section those essential ingredients of the original and complementary ADHM linear sigma models that will be part of the final model. These can be introduced either following the lines in the original approach of Ref.\cite{Witten:1994tz} or the one adopted in \cite{Lambert:1995dp}. In Ref. \cite{Ali:2023csc} we used the latter while in the present we shall  employ the former. 

First we shall collect ingredients from Witten's original model.

In this case there are bosons $X^{AY}$ and $\phi^{A'Y'}$ with $A$, $A'$=1,2, $Y = 1, 2,..., 2k$ and $Y' = 1,2,..., 2k'$. The $A$ and $A'$ indices are raised (lowered) with the antisymmetric $Sp(1)$ tensors $\epsilon^{AB}$($\epsilon_{AB}$) and $\epsilon^{A'B'}$($\epsilon_{A'B'}$) respectively. Similarly the $Y$ and $Y'$ indices are raised(lowered) with the help of the antisymmetric $Sp(k)$ and $Sp(k')$ tensors $\epsilon^{YZ}$($\epsilon_{YZ}$) and $\epsilon^{Y'Z'}$($\epsilon_{Y'Z'}$) respectively. The light cone coordinates $\sigma^\pm$ are defined as $\sigma^\pm=(\tau\pm\sigma)/
\sqrt 2$ and $\partial_{\pm} = \frac{1}{\sqrt{2}}(\partial_0 \pm \partial_1)$  while the  worldsheet metric is taken as $ds^2=d\tau^2-d\sigma^2$. The fermionic partners of $X$ and $\phi$ bosons are $\psi_-^{A'Y}$ and $\chi_-^{AY'}$ respectively and corresponding supersymmetry transformations are 
\begin{eqnarray}\label{susytransf}
\delta_{\eta} X^{AY}&=&i \epsilon_{A'B'}\eta^{AA'}_{+}\psi^{B'Y}_{-}, ~~
\delta_{\eta} \psi^{A'Y}_{-} = \epsilon_{AB}\eta^{AA'}_{+} \partial_{-} X^{BY},\nonumber \\ 
\delta_{\eta} \phi^{A'Y'}&=&i \epsilon_{AB}\eta^{AA'}_{+}{\chi}^{BY'}_{-}, ~~~
\delta_{\eta}{\chi}^{AY'}_{-} = \epsilon_{A'B'}\eta^{AA'}_{+} \partial_{-} \phi^{B'Y'}.
\end{eqnarray}
There are supercharges $Q^{AA'}$ that obey the superalgebra 
\begin{equation}
\label{susyalgebra}
\{Q^{AA'},Q^{BB'}\}=\epsilon^{AB}\epsilon^{A'B'}
P^+,~~~P^+=P_-=-i\partial/\partial \sigma^-.
\end{equation}
The fields and supercharges have the following reality conditions
\begin{equation}
X^{A Y}=\epsilon^{AB}\epsilon^{YZ}\overline
X_{BZ},~~~	\phi^{A'Y'}=\epsilon^{A'B'}\epsilon^{Y'Z'}\overline \phi_{B'Z'},~~~Q^{AA'}=\epsilon^{AB}\epsilon^{A'B'}
Q^\dagger_{BB'}.
\label{realcon}	
\end{equation}
Additionally there are left moving fermions $\lambda_+^a$, $a=1,\dots, n$ with the action
\begin{equation}
\label{partof}
\frac{i}{2}\int d^2\sigma\left(\lambda_+^a\partial_-\lambda_+^a
-\lambda_+^aG_{a\theta}\rho_-^\theta\right),
\end{equation}
where the factor $\rho_-^\theta$ in the Yukawa term contains both $\psi$'s and $\chi$'s. The expression for $G_{a\theta}$ is fixed using the equation of motion of
$\lambda_+^a$
\begin{equation}
\label{artof}
\partial_-\lambda_+^a=G^a_\theta\rho_-^\theta
\end{equation}
such that the most general supersymmetry transformation of $\lambda_+^a$
\begin{equation}
\label{susylamda}\delta\lambda_+^a= \eta_+^{AA'}C_{AA'}^a,
\end{equation}
is in consonance with supersymmetry algebra (\ref{susyalgebra}) above. Eqn.(\ref{susylamda}) gives
\begin{eqnarray}
\label{deletapeta}
\delta_{\eta'}\delta_\eta \lambda_+^a =i\eta_+^{AA'} \left(\frac{\partial C_{AA'}^a}{\partial X^{BY}}\epsilon_{B'C'} {\eta'}_+^{BB'} \psi_{-}^{C'Y} +\frac{\partial C_{AA'}^a	}{\partial\phi^{B'Y'}} \epsilon_{BC} {\eta'}_+^{BB'}\chi^{CY'}_{-} \right),
\end{eqnarray}
such that
\begin{eqnarray}
	[\delta_{\eta'},\delta_\eta]
	= -i\epsilon_{AB}\epsilon_{A'B'}
	\eta_+^{AA'}\eta'{}_+^{BB'}\partial_-\lambda_+^a \cr
  	=-i\epsilon_{AB}\epsilon_{A'B'} \eta_+^{AA'}\eta'{}_+^{ BB'}G^a_\theta\rho^\theta. 
	\label{commu}
\end{eqnarray}

Eqn.(\ref{deletapeta}) conforms with (\ref{commu}) provided
\begin{eqnarray}
\label{adhmcondc}
  \frac{\partial C_{AA'}^a}{\partial X^{B Y}}
+\frac{\partial C_{BA'}^a}{\partial X^{A Y}}
=  \frac{\partial C_{AA'}^a}{\partial \phi^{B'Y'}}
+\frac{\partial C_{AB'}^a}{\partial \phi^{A'Y'}}=0.
\end{eqnarray}
To obtain the Lagrangian with above symmetry $\lambda^a$ is taken to be a part of a supermultiplet with
\begin{equation}
\label{miccon}
\Lambda^a=\lambda^a+\theta F^a,~~~ \delta\lambda^a=\eta F^a
\end{equation}
such that potential becomes
\begin{equation}
\label{iccon}
V_W=\frac{1}{2}\sum_aF^aF^a=\frac{1}{2}\sum_a C^aC^a
\end{equation}
with the definition and condition where $C^a$ are defined using $c$-numbers $c^{AA'}$ with following normalization condition

\begin{equation}
\label{uj}
C^a=c^{AA'}C_{AA'}^a, ~~~\epsilon_{AB}\epsilon_{A'B'}c^{AA'}c^{BB'}=1.
\end{equation}
The condition for the Lagrangian to be invariant under all supersymmetries is
\begin{equation}
\label{caacond}
\sum_a\left(C_{AA'}^aC_{BB'}^a+C_{BA'}^aC_{AB'}^a
\right)=0.
\end{equation}

These are the ingredients from the original model. 

Then there are corresponding ingredients from Ali-Ilahi's  complementary model. 

This sector has $X\rightarrow \hat{X}$, $\phi\rightarrow\hat{\phi}$, $\lambda^{a} \rightarrow  {\hat\lambda}^{\hat{a}}$, $\hat{a}=1,...,n'$, $G^{a}_{\theta} \rightarrow  \hat{G}^{\hat{a}}_{\theta}$, $C^{a}_{AA'} \rightarrow \hat {C}^{\hat{a}}_{AA'}$, $\Lambda^{a} \rightarrow \hat {\Lambda}^{\hat{a}}$, $F^{a} \rightarrow \hat {F}^{\hat{a}}$ and $c^{AA'} \rightarrow \hat c^{AA'}$ with corresponding equations duplicating Eqs.(\ref{partof}-\ref{caacond}). 
The full Lagrangian is obtained by judicious combination of elements from original and complementary models. It is
\begin{equation}
L=L_{kin}+L_W+L_C
\end{equation}
where $L_W$ is
\begin{eqnarray}
\label{witten2.21} 
L_W = - \frac{i}{4} \int d^2 \sigma \lambda_+^a \left(\epsilon^{BD}\frac{\partial C^a_{BB'}}{\partial X^{DY}}\psi_-^{B'Y} +\epsilon^{B'D'}\frac{\partial C^a_{BB'} }{\partial \phi^{D'Y'}}\chi_-^{BY'} \right) \cr  -\frac{1}{ 8}\int d^2\sigma \epsilon^{AB}\epsilon^{A'B'}C^a_{AA'}C^a_{BB'}
\end{eqnarray}
with $C^a_{AA'}$  given by Eqn.(\ref{witten2.23}).
\begin{eqnarray}
L_C = - \frac{i}{ 4} \int d^2 \sigma \hat \lambda_+^a \left(\epsilon^{BD}\frac{\partial \hat C^a_{BB'}}{\partial \hat{X}^{DY}}\hat\psi_-^{B'Y} +\epsilon^{B'D'}\frac{\partial \hat C^a_{BB'}}{\partial \hat\phi^{D'Y'}}\hat\chi_-^{BY'} \right) \cr  -{1}{ 8}\int d^2\sigma \epsilon^{AB}\epsilon^{A'B'}\hat C^a_{AA'} \hat C^a_{BB'}.
\end{eqnarray}
To obtain explicit form of $C_{AA'}^a$, in Witten's original model,  we begin with the following general form that is separately linear in both $\phi$ and $X$.
\begin{eqnarray}\label{witten2.22}
C_{AA'}^a&=M_{AA'}^a+X_{AY}N^{a\,Y}_{A'}
+\phi_{A'}{}^{Y'}D_{AY'}^a +X_{A}{}^{Y}\phi_{A'}{}^{Y'}E^a_{YY'}.
\end{eqnarray}

Then we make the choice is to take $M=N=0$ such that 
\begin{eqnarray}\label{witten2.23}
{C}^{a}_{AA'}=\epsilon_{A'B'}D^{a}_{AY'}\phi^{~B'Y'}+ \epsilon_{AB}\epsilon_{A'B'}E^{a}_{YY'}X^{BY}\phi^{~B'Y'} \equiv \phi_{A'}^{~Y'}B^{a}_{AY'}(X)
\end{eqnarray}
such that the condition (\ref{caacond}) reduces to

\begin{equation}\label{B}
\sum_{a}(B^{a}_{AY'}(X)B^{a}_{BZ'}(X)+B^{a}_{BY'}(X)B^{a}_{AZ'}(X))=0.
\end{equation}
Here $B^{a}_{AY'}(X)$ is linear in $X$ and independent of $\phi$.

For the original model there are $N=n-4k'$ massless components with $v_i^a(=v^a_i(X), i=1,2,...,N$ and
\begin{equation}\label{mlessv}
\sum_{a}v^{a}_{i}(X)B^{a}_{AY'}(X)=0.
\end{equation}
The normalization condition is 
\begin{equation}\label{orthonorm}
\sum_{a}v^{a}_{i}(X)v^{a}_{j}(X)=\delta_{ij}.
\end{equation}
Setting 
\begin{equation}\label{setlamb}
\lambda^{a}_{+}=\sum_{i=1}^{N}v^{a}_{i}(X)\lambda_{+i}
\end{equation}
amount to putting massive modes of $\lambda^{a}_{+}$ equal to zero.

The covariant derivative and the resulting instanton field are
\begin{equation}\label{covderinst}
\delta_{ij}\partial_{-}+ A_{ijAY}\partial_{-}X^{AY},~~~A_{ijAY}=\sum_{a}v^{a}_{i}(X)\frac{\partial v^{a}_{j}(X)}{\partial X^{AY}}
\end{equation}
respectively.
The supersymmetry structure so far is the usual $(0,4)$ one.

This is Witten's $k'$ instanton in ${\mathcal R}^{4k}$.

Equal amount of construction is borrowed from the complementary sigma model. Which has 
the following general form of the tensor $\hat{C}^{\hat a}_{AA'}$
\begin{eqnarray}\label{genchat}
\hat C_{AA'}^{\hat{a}}&=\hat M_{AA'}^{\hat{a}}+X_{AY}\hat N^{\hat a\,Y}_{A'}
+\phi_{A'}{}^{Y'}\hat D_{AY'}^{\hat{a}} +X_{A}{}^{Y}\phi_{A'}{}^{Y'}\hat E^{\hat{a}}_{YY'}.
\end{eqnarray}

In the complementary model we choose $\hat M=\hat D=0$ such that the result is 
\begin{eqnarray}\label{caphi}
\hat{C}^{\hat a}_{AA'}=\epsilon_{AB} \hat{N}^{\hat a}_{A'Y}X^{BY}+ \epsilon_{AB}\epsilon_{A'B'}\hat{E}^{\hat a}_{YY'}\hat{X}^{BY}\hat\phi^{~B'Y'}\equiv X_{A}^{~Y}A^{\hat a}_{A'Y}(\hat\phi).
\end{eqnarray}
The condition corresponding to Eqn.(\ref{caacond})
\begin{equation}
\sum_{\hat{a}}\left(\hat C_{AA'}^{\hat{a}}\hat C_{BB'}^{\hat{a}}+\hat C_{BA'}^{\hat{a}} \hat C_{AB'}^{\hat{a}}
\right)=0
\end{equation}
now becomes
\begin{equation}\label{aaacond}
\sum_{\hat{a}}(A^{\hat a}_{A'Y}(\hat\phi)A^{\hat a}_{B'Z}(\hat\phi)+A^{\hat a}_{B'Y}(\hat\phi)A^{\hat a}_{A'Z}(\hat\phi))=0.
\end{equation}
 Here $A^{\hat a}_{A'Y}(\phi)$ is linear in $\hat\phi$ and independent of $\hat{X}$.

For complementary model there are $N'=n'-4k$ massless components $\hat v_{\hat i}^{\hat a}(\hat\phi), \hat i=1,2,...,N'$ with
\begin{equation}\label{mlessv2}
\sum_{\hat a}(\hat\phi)\hat v^{\hat a}_{\hat i}A^{\hat a}_{A'Y}(\hat\phi)=0.
\end{equation}
For the complementary part the normalization condition will be
\begin{equation}\label{orthonor2}
\sum_{\hat a}\hat v^{\hat a}_{\hat i}(\hat\phi)\hat v^{\hat a}_{\hat j}(\hat\phi)=\delta_{\hat{i} \hat{j}}.
\end{equation}
Eqn. corresponding to (\ref{setlamb}) will be
\begin{equation}\label{setlamb2}
\hat \lambda^{\hat a}_{+}=\sum_{\hat i=1}^{N'}\hat v^{\hat a}_{\hat i}(\hat\phi)\hat \lambda_{+ \hat i}
\end{equation}
that amounts to putting the massive fermions to zero.

In the complementary case the covariant derivative has the expression
\begin{equation}\label{covderexp}
\delta_{\hat i \hat j}\partial_{-}+ A_{\hat i \hat jA'Y'}\partial_{-}\phi^{A'Y'}
\end{equation}
with the instanton
\begin{equation}\label{inst}
\hat{A}_{\hat i \hat jA'Y'}=\sum_{\hat a}\hat v^{\hat a}_{\hat i}(\hat\phi)\frac{\partial \hat v^{\hat a}_{\hat j}(\hat\phi)}{\partial \phi^{A'Y'}}.
\end{equation}

This is Ali-Ilahi's $k$ instanton in ${\mathcal R}^{4k'}$.

In combination we get a $k$ instanton plus a  $k'$ instanton in ${\mathcal R}^{4k+4k'}$.

It is clear from this discussion that while the minimal original and
minimal Ali-Ilahi models are defined in ${\mathcal R}^4$ the minimal complete model is defined in ${\mathcal R}^8$.

This completes our construction of the complete ADHM instanton sigma model by amalgamation of Witten's original and Ali-Ilahi's complementary model.

With these additional Yukawa couplings the  supersymmetry structure should be a generalization of the $(0,4)$ supersymmetry and we would term it as large $(0,4)$. Here we are anticipating this new supersymmetry structure with analogy to the superconformal case. In our knowledge corresponding algebra has not been worked out.

\section{The 1+1 Instanton Case}\label{one}

To analyze the structure of the the moduli space in case of the original plus the complementary models, that is the complete model, it is convenient to go to $(1,1)$ instanton cases. 
Here the single instantons from both original and complementary branches are present.

When we  put $k=1$ in the former case and $k'=1$ in the latter case to get $R^4$ spacetime from both cases we realize that the minimal non-trivial construction lives on $R^8$. The resulting models have $F' \times K  \times SU(2)_L \times SU(2)_R \cong SU(2)^4$ invariance in the original model sector and $F\times K' \times SU(2)'_L \times SU(2)'_R \cong SU(2)^4$ in the complementary model sector.

The resulting left moving massless fermions in the original model are  $\lambda_{+}^{AY'}$ and $\lambda_{+}^{YY'}$  with the reality conditions  
\begin{eqnarray}\label{rcondlamb1}
\lambda^{AY'}_{+}= \epsilon^{AB}\epsilon^{Y'Z'} \overline {\lambda}_{+BZ'},~&&     \lambda^{YY'}_{+}= \epsilon^{YZ}\epsilon^{Y'Z'} \overline {\lambda}_{+ZZ'}
\end{eqnarray} 
 while in the complementary case these are $\hat \lambda_{+}^{A'Y}$ and $\hat \lambda_{+}^{YY'}$ with reality conditions
 \begin{eqnarray}\label{rcondlam2}
\hat \lambda^{A'Y}_{+}= \epsilon^{A'B'}\epsilon^{YZ} \overline {\hat \lambda}_{+B'Z},~&&     \hat \lambda^{YY'}_{+}= \epsilon^{YZ}\epsilon^{Y'Z'} \overline {\hat \lambda}_{+ZZ'}.
 \end{eqnarray}
 The Yukawa couplings of the original model are 
 \begin{eqnarray}\label{yukcoup1}
 {C}^{YY'}_{~~~~BB'}= X_{B}^{~Y} \phi^{~Y'}_{B'},~ && 
 {C}^{AY'}_{~~~~BB'}= \frac{\rho}{\sqrt{2}} \delta^{A}_{~B}\phi^{~Y'}_{B'}  
 \end{eqnarray}
 and of the complementary model are as follows
 \begin{eqnarray}\label{yukcoup2}
\hat {C}^{YY'}_{~~~~BB'}= X_{B}^{~Y} \phi^{~Y'}_{B'},~ && 
\hat {C}^{A'Y}_{~~~~BB'}= \frac{\omega}{\sqrt{2}} \delta^{A'}_{~B'}X^{~Y}_{B}.  
 \end{eqnarray}
 The resulting potentials for Witten's original and Ali-Ilahi's complementary model are
 \begin{equation}\label{potentials}
 V_W = \frac{1}{8}(X^{2}+\rho^{2})\phi^{2},~~~	V_C = \frac{1}{8}(\phi^{2}+\omega^{2})X^{2}	
 \end{equation}
 respectively. Using the fermionic field ans\"atz
 \begin{eqnarray}\label{ffanst1}
 \lambda^{YY'} = \frac{\rho\zeta^{~YY'}_+}{\sqrt{\rho^2 + X^2}},~ && \lambda^{AY'} = -\frac{\sqrt{2}X^{A}_{~~Y}\zeta_{+}^{~YY'}}{\sqrt{\rho^2 + \phi^2}}.
 \end{eqnarray}
 for the original case and 
 \begin{eqnarray}\label{ffanst2}
 \hat \lambda^{YY'} = \frac{\omega\zeta^{~YY'}_+}{\sqrt{\omega^2 + \phi^2}},~ && \hat \lambda^{A'Y} = -\frac{\sqrt{2}\phi^{A'}_{~~Y'}\zeta_{+}^{~YY'}}{\sqrt{\omega^2 + \phi^2}}
 \end{eqnarray} 
 for the complementary case, the instanton expressions take the forms
 \begin{eqnarray}\label{instexp1}
 &&\lambda_{+YY'}\partial_{-}\lambda^{YY'}_{+}+\lambda_{+AY'}\partial_{-}\lambda^{AY'}_{+} = \zeta_{+YY'}\partial_{-}\zeta^{~YY'}_{+} -\zeta_{+YY'}\nonumber\\
 &&\frac{\frac{1}{2}\epsilon_{AB}(X^{AY}\partial_{-}X^{BZ}+X^{AZ}\partial_{-}X^{BY})}{X^2+\rho^2}\zeta^{~~~Y'}_{+Z}
 \end{eqnarray}
 and 
 \begin{eqnarray}\label{instexp2}
 &&\hat \lambda_{+YY'}\partial_{-}\hat \lambda^{YY'}_{+}+\hat \lambda_{+A'Y}\partial_{-}\hat \lambda^{A'Y}_{+} = \hat \zeta_{+YY'}\partial_{-}\hat \zeta^{~YY'}_{+} -\hat \zeta_{+YY'}\nonumber\\&&\frac{\frac{1}{2}\epsilon_{A'B'}(\phi^{A'Y'}\partial_{-}\phi^{B'Z'}+\phi^{A'Z'}\partial_{-}\phi^{B'Y'})}{\phi^2+\omega^2}\hat \zeta^{Y}_{+Z'} 
 \end{eqnarray}
 for original and complementary branches respectively.

The total potential in case of double instanton is
\begin{equation}\label{totpot}
V=V_W+V_C=\frac{1}{8}\left(2X^2\phi^2+\rho^2\phi^2+\omega^2X^2\right).
\end{equation}

A single look at the potential (\ref{totpot}) reveals that in case of the complete ADHM instanton sigma model there is no moduli space. If the potential for one instanton vanishes then the potential of the other is non-zero and vice versa.

\section{Discussion}\label{discussion}

The duality between the original and complementary constructions becomes a $Z_2$ symmetry in the complete model. This is in consonance with what we expect from the infrared limit of the model.

In the infrared limit, or rather the conformal limit, both original and complementary models will have small N=4 superconformal symmetries \cite{Ademollo:1975an, Ademollo:1976pp}. The complete model will have large N=4 superconformal symmetry \cite{Sevrin:1988ab, Ivanov:1988rt, Ali:2000we, Ali:2000zu, Ali:2003aa} in the conformal limit. We need better understanding of all of these limits.

The small N=4 superconformal symmetry has an $SU(2)$ Kac-Moody subalgebra. This should be visible as an $SU(2)$ R symmetry before the conformal limit. The original as well as the complementary models have a galore of $SU(2)$ symmetries before the conformal limit. The supersymmetry structure in the present formulation  is not entirely transparent.  Construction of the complementary and complete 't Hooft instantons, taking Callan-Harvey-Strominger as the starting original construction, should help in clarifying these issues because the structure of supersymmetry is much more clear in that case. These constructions should also help in clarifying the semi-infinite throat there and its counter part here. It would also be interesting to analyze in what limit the ADHM construction and its generalizations to higher groups reduce to 't Hooft construction and corresponding generalizations.

The complete model has two $SU(2)$ R symmetries. The large N=4 superconformal algebra has an $SU(2) \times SU(2) \times U(1)$ Kac-Moody subalgebra and hence at least in the superconformal limit we should be able to identify a $U(1)$ symmetry in the complete model discussed in this note. Our speculation in this regard is the following. In the pair of scalars $(X, \phi)$ we have two linear combinations and one of them is $U(1)$  field while another one is a scalar field with background charge. Clearly these aspects need further clarifications. It will also be interesting to do In\"on\"u-Wigner contraction of the complete model and reach the middle N=4 superconformal symmetry and corresponding ADHM linear sigma model \cite{Ali:1993sd,Ali:2000we,Hasiewicz:1989vp}. 

From the resulting structure it is apparent that the quantization of the original model discussed in Ref.\cite{Lambert:1995dp} and the quantization of the complementary model discussed in Ref.\cite{Ali:2023zxt} cover all that is needed for quantization of this model.

In Ref.\cite{Papadopoulos:2024uvi} Papdopoulos and Witten gave a direct proof of the fact that in two dimensions scale invariance implies conformal invariance. A related problem is to analyze the flow of the model considered in this note in the infrared and see how it flows to large $N=4$ superconformal symmetry. Also we would like to know the ADHM sigma model that flows to middle $N=4$ superconformal algebra \cite{Ali:2024amc}.

Also connection of the present construction to AdS/CFT correspondence \cite{Maldacena:1997re, Witten:1998qj, Gubser:1998bc} should be probed more deeply. In a recent development off-shall harmonic superspace formalism for the complementary ADHM instanton sigma model has been discussed in the papers \cite{Ali:2025jcu} and \cite{Ali:2025ntc}.

\textit{Acknowledgements}: We thank Dr. Mohsin Ilahi and Dr. Shafeeq Rahman Thottoli for discussions. This work was done as part of the Ph.D. thesis of P.P. Abdul Salih. We thank Prof. Neil D. Lambert for some clarifications and Dr. Raktim Abir for a very useful suggestion.


\begin{thebibliography}{999}
\bibitem{Callan:1991dj}
C.~G.~Callan, Jr., J.~A.~Harvey and A.~Strominger,
``World sheet approach to heterotic instantons and solitons,''
Nucl.\ Phys.\ B {\bf 359} (1991) 611.
	
\bibitem{Callan:1991ky}
C.~G.~Callan, Jr., J.~A.~Harvey and A.~Strominger,
``Worldbrane actions for string solitons,''
Nucl.\ Phys.\ B {\bf 367} (1991) 60.
	
\bibitem{Callan:1991at}
C.~G.~Callan, Jr., J.~A.~Harvey and A.~Strominger,
``Supersymmetric string solitons,'' in: Proc. String Theory and Quantum Gravity '91(Trieste, 1991)
[arXiv:hep-th/9112030].
	
		
\bibitem{Atiyah:1978ri}
M.~F.~Atiyah, N.~J.~Hitchin, V.~G.~Drinfeld and Y.~I.~Manin,
``Construction of Instantons,''
Phys.\ Lett.\ A {\bf 65} (1978) 185.
		
\bibitem{Witten:1978qe}
E.~Witten,
``Some Comments on the Recent Twistor Space Constructions,'' Workshop on Applications of Complex Manifold Techniques to Problems in Theoretical Physics, 1978
	
\bibitem{Corrigan:1983sv}
E.~Corrigan and P.~Goddard,
``Construction of Instanton and Monopole Solutions and Reciprocity,''
Annals Phys. \textbf{154} (1984), 253
	
\bibitem{Christ:1978jy}
N.~H.~Christ, E.~J.~Weinberg and N.~K.~Stanton,
``General self-dual Yang-Mills Solutions,''
Phys. Rev. D \textbf{18} (1978), 2013

\bibitem{Corrigan:1978ce}
E.~Corrigan, D.~B.~Fairlie, S.~Templeton and P.~Goddard,
``A Green's Function for the General Selfdual Gauge Field,''
Nucl. Phys. B \textbf{140}, 31-44 (1978)
	
\bibitem{Donaldson:2022}S. Donaldson, ``The ADHM construction of Yang-Mills instantons",  arXiv:2205.08639v1 [math.DG]	
	
	
\bibitem{Witten:1994tz}
E.~Witten,
``Sigma models and the ADHM construction of instantons,''
J.\ Geom.\ Phys.\  {\bf 15} (1995) 215
[hep-th/9410052].
	 
	
		
\bibitem{Lambert:1995dp}
N.~D.~Lambert,
``Quantizing the (0,4) Supersymmetric ADHM Sigma Model,''
Nucl. Phys. B \textbf{460} (1996), 221-232
[arXiv:hep-th/9508039].

\bibitem{Douglas:1996uz}
M.~R.~Douglas,
``Gauge fields and D-branes",
J. Geom. Phys. \textbf{28} (1998), 255-262
[arXiv:hep-th/9604198].

\bibitem{Lambert:1996yd}
N.~D.~Lambert,
``Heterotic p-branes from massive sigma models,''
Nucl. Phys. B \textbf{477}, 141-154 (1996)
[arXiv:hep-th/9605010].
	
\bibitem{Lambert:1997gs}
N.~D.~Lambert,
``D-brane bound states and the generalized ADHM construction,''
Nucl. Phys. B \textbf{519}, 214-224 (1998)
[arXiv:hep-th/9707156].
	
\bibitem{Johnson:1998yw}
C.~V.~Johnson,
``On the (0,4) conformal field theory of the throat,''
Mod. Phys. Lett. A \textbf{13}, 2463-2474 (1998)
[arXiv:hep-th/9804201].
		
\bibitem{Ali:2023csc}
A.~Ali and M.~Ilahi,
``Complementary ADHM Instanton Sigma Model,''
[arXiv:2305.05951 [hep-th]].

\bibitem{Ali:2023kkf}
A.~Ali and M.~Ilahi,
``$Z_2$ Symmetry of $AdS_3 \times S^3 \times S^3 \times S^1$ Superstrings,''
[arXiv:2306.13970 [hep-th]].

\bibitem{Ali:2023xov}
A.~Ali,
``Superalgebra Doubling in $AdS_3$ Superstrings,''
[arXiv:2306.11047 [hep-th]].
	
\bibitem{Witten:1995zh} E. Witten, ``Some Comments on String Dynamics" [hep-th/9507121], in Strings '95, Future Perspectives in String Theory, ed. I. Bars et. al.
		
\bibitem{Ali:2023zxt}
A.~Ali, M.~Ilahi and S.~R.~Thottoli,
``Quantization of Complementary ADHM Sigma Model,''
[arXiv:2306.10002 [hep-th]].
		
\bibitem{Giveon:1998ns}
A.~Giveon, D.~Kutasov and N.~Seiberg,
``Comments on String Theory on $AdS_3$,''
Adv.\ Theor.\ Math.\ Phys.\  {\bf 2} (1998) 733
[hep-th/9806194].


\bibitem{Banados:1992wn}
M.~Banados, C.~Teitelboim and J.~Zanelli,
``The Black hole in three-dimensional space-time,''
Phys. Rev. Lett. \textbf{69}, 1849-1851 (1992)
[arXiv:hep-th/9204099 [hep-th]].

\bibitem{Ali:1992mj}
A.~Ali and A.~Kumar,
``O (d, d) transformations and 3-D black hole,''
Mod. Phys. Lett. A \textbf{8}, 2045-2052 (1993)
[arXiv:hep-th/9303032 [hep-th]].
	
\bibitem{Ademollo:1975an}
M.~Ademollo, L.~Brink, A.~D'Adda, R.~D'Auria, E.~Napolitano, S.~Sciuto, E.~D.~Giudice, P.~D.~Vecchia, S.~Ferrara, F.~Gliozzi, R.~Musto and R.~Pettorino,
``Supersymmetric Strings and Color Confinement,''
Phys.\ Lett.\ B {\bf 62} (1976) 105.
	
\bibitem{Ademollo:1976pp}
M.~Ademollo, L.~Brink, A.~D'Adda, R.~D'Auria, E.~Napolitano, S.~Sciuto, E.~D.~Giudice, P.~D.~Vecchia, S.~Ferrara, F.~Gliozzi, R.~Musto, R.~Pettorino and J.~Schwarz,
``Dual String with $U(1)$ Color Symmetry,''
Nucl.\ Phys.\ B {\bf 111} (1976) 77.

\bibitem{Sevrin:1988ab}
A. Sevrin, W. Troost and A. van Proeyen, ``Superconformal Algebras in Two-Dimensions with N=4'', Phys. Lett. B 208 (1988) 447.
	 
\bibitem{Ivanov:1988rt}
E.~A.~Ivanov, S.~O.~Krivonos and V.~M.~Leviant,
``Quantum N=3, N=4 Superconformal WZW Sigma Models,''
Phys.\ Lett.\ B {\bf 215} (1988) 689
Erratum: [Phys.\ Lett.\ B {\bf 221} (1989) 432]
		
\bibitem{Ali:2000we}
A.~Ali,
``Conformal Symmetry of Superstrings on $AdS_3 \times S^3 \times T^4$ and D1 / D5 system,''
Mod.\ Phys.\ Lett.\ A {\bf 17} (2002) 2477
[hep-th/0007021].
		
\bibitem{Ali:2000zu}
A.~Ali,
``Free Field Realizations of N=4 Superconformal Algebras,''
Indian J. Pure Appl. Phys. \textbf{38} (2000), 446-452
		
\bibitem{Ali:2003aa}
A.~Ali,
``Types of Two-dimensional N = 4 Superconformal Field Theories,''
Pramana \textbf{61} (2003), 1065-1078
[arXiv:hep-th/9906096].
	
 	
\bibitem{Ali:1993sd}
A.~Ali and A.~Kumar,
``A New N=4 Superconformal Algebra,''
Mod. Phys. Lett. A \textbf{8}, 1527-1532 (1993)
[arXiv:hep-th/9301010].
	
\bibitem{Hasiewicz:1989vp}
Z.~Hasiewicz, K.~Thielemans and W.~Troost,
``Superconformal Algebras and Clifford Algebras,''
J. Math. Phys. \textbf{31}, 744 (1990)

\bibitem{Papadopoulos:2024uvi}
G.~Papadopoulos and E.~Witten,
``Scale and Conformal Invariance in 2d Sigma Models, with an Application to N=4 Supersymmetry,''
[arXiv:2404.19526 [hep-th]].

\bibitem{Ali:2024amc}
A.~Ali, M.~Ilahi, P.~P.~A.~Salih and S.~R.~Thottoli,
``Symmetry Structure of ADHM Sigma Models and $AdS_3$ Superstrings,''
[arXiv:2409.00474 [hep-th]].

\bibitem{Maldacena:1997re}
J.~M.~Maldacena,
``The Large N limit of superconformal field theories and supergravity,''
Adv. Theor. Math. Phys. \textbf{2}, 231-252 (1998)
[arXiv:hep-th/9711200 [hep-th]].

\bibitem{Witten:1998qj}
E.~Witten,
``Anti-de Sitter space and holography,''
Adv. Theor. Math. Phys. \textbf{2}, 253-291 (1998)
[arXiv:hep-th/9802150 [hep-th]].

\bibitem{Gubser:1998bc}
S.~S.~Gubser, I.~R.~Klebanov and A.~M.~Polyakov,
``Gauge theory correlators from noncritical string theory,''
Phys. Lett. B \textbf{428}, 105-114 (1998)
[arXiv:hep-th/9802109 [hep-th]].

\bibitem{Ali:2025ntc}
A.~Ali, M.~Ilahi, P.~P.~A.~Salih and S.~R.~Thottoli,
``Harmonic Superspace for Ali-Ilahi's ADHM Instanton Sigma Model,''
[arXiv:2507.22948 [hep-th]].

\bibitem{Ali:2025jcu}
A.~Ali, M.~Ilahi, P.~P.~A.~Salih and S.~R.~Thottoli,
``Off-shell Formalism for Ali-Ilahi's ADHM Instanton Sigma Model,''
[arXiv:2507.11305 [hep-th]].
	
\end{thebibliography}

\end{document}